# Dynamics of Vesicle Formation from Lipid Droplets: Mechanism and Controllability

Zilu Wang and Xuehao He*

Department of Polymer Science and Engineering, School of Chemical Engineering and Technology, Tianjin University, 300072 Tianjin, China

ABSTRACT: A coarse-grained model developed by Marrink et al. (J. Phys. Chem. B 111, 7812, 2007) is applied to investigate vesiculation of lipid (DPPC) droplets in water. Three kinds of morphologies of micelles are found with increasing lipid droplet size. When the initial lipid droplet is smaller, the equilibrium structure of the droplet is a spherical micelle. When the initial lipid droplet is larger, the lipid ball starts to transform into a disk micelle or vesicle. The mechanism of vesicle formation from a lipid ball is analyzed from the self-assembly of DPPC on the molecular level, and the morphological transition from disk to vesicle with increasing droplet size is demonstrated. Importantly, we discover that the transition point is not very sharp, and for a fixed-size lipid ball, the disk and vesicle appear with certain probabilities. The splitting phenomenon, i.e., the formation of a disk/vesicle structure from a lipid droplet, is explained by applying a hybrid model of Helfrich membrane theory. The elastic module of the DPPC bilayer and the smallest size of a lipid droplet for certain formation of a vesicle are successfully predicted.

*Corresponding author, E-mail address: xhhe@tju.edu.cn

# 1. INTRODUCTION

Lipid molecules, typically amphiphilic molecules with dual structures exhibiting different interactions with solvents, can spontaneously self-assemble into various ordered structures, such as spherical, rod-like, and lamellar micelles.[1,2] Among these structures, the vesicle is a special lamellar structure that forms closed compartments playing important functions in intracellular matter transportation, endo- and exocytosis, drug delivery, and biomolecular engineering.[3-6] The vesicle transition has been a hot topic in research on the vesiculation of amphiphilic molecules.[14-26] The formation of lipid vesicles depends not only on molecular structure, but also on external conditions such as heat and light,[7,8] metal-cations,[9,10] pH,[11] proteins[12], and shearing.[13] Modern experimental technologies, such as time-resolved X-ray scattering and real-time observation with microscopes,[26-28] have been applied to study the micellization of lipids or amphiphilic molecules. However, dynamic information related to the molecular details of vesicle formation is difficult to obtain directly. Instead of experimental methods, computer molecular simulation has the capability to provide more detailed information, and has become a powerful tool in the study of the mechanism of molecular micellization.[29] Many simulation methods, including the Brownian Dynamic method (BD),[30] Molecular Dynamics (MD),[31] and Dissipation Particle Dynamics (DPD)[32,33], have been applied to study different problems in the self-assembly of lipids. The earliest work regarding vesicle formation was carried out by Drouffe et al.,[34] in which a lipid molecule was represented by an anisotropic single particle. Since then, simulation work[12,31,33,35-50] related to the micellization of amphiphilic molecules,[33,35,36] the creation and fusion of vesicles[31,37-41], the phase behaviors and structures of single or mixed surfactants[42,43], membrane-protein interactions[12], interface or surface properties[44], and bilayer properties[43,45-50] have also been carried out.

It is noteworthy that most simulation work in the past has focused on the aggregation of lipid molecules starting with randomly distributed lipids in a solution. The formation of vesicles in such a case strongly depends on the density of lipids in the lipid solution and the diffusion of molecules. Usually, a wide spectrum of vesicles of various sizes accompanying small lipid clusters are obtained; this has been confirmed in many experiments. Preparation of vesicles with controllable, uniform size and structure is necessary to design for specific functionality in applications of advanced release systems and material templates. Whether we can tune vesicle size by controlling the number of aggregating lipids is an important question. In this paper, we carried out a simulation of vesicle formation, starting with an initial state of a lipid droplet (or lipid ball), to explore the dependence of vesicle formation on the size of the lipid ball and the self-assembly behavior of lipid molecules. After that, a simple hybrid model of Helfrich's membrane elastic theory[23,24,51-54] is proposed to explain the phenomenon of splitting of morphologies (vesicle/disk) from a lipid droplet when the initial number of lipid molecules in the droplet approaches the transition value from disk to vesicle. The simulation results and molecular information obtained will help to provide a deeper understanding of the vesiculation mechanism of lipids and the relationship between the metastability of vesicle structures and size.

## 2. MOLECULAR DYNAMICS SIMULATION

A coarse-grained model developed by Marrink et al. with well-optimized force field parameters (MARTINI force field V2.0)[55] for high efficiency and reliable thermodynamic properties[55-57] is applied to simulate the varied micellizations of lipids from lipid droplets. In this model, each of the four major atoms in a lipid are mapped to one coarse-grained (CG) bead, and one CG water bead represents four real water molecules. The differences in interaction between CG beads for nonbonded interactions are described with the Lenard-Jones potential, and reflect their hydrophilic and hydrophobic properties. We selected the Dipalmitoylphosphatidylcholine (DPPC) molecule as the model molecule in our simulation. A DPPC molecule can be described with 12 CG beads (see Fig. 1). Compared with an atomistic simulation, the effective time factor in this simulation is 4, which means the simulation time is 4 times as fast as the time we set. This accelerating factor is mainly due to the smoother energy landscape in this CG model with less friction between particles.[55]

All simulations in this work are carried out in two steps: first, preparing the lipid droplet; then, simulating vesicle formation. First, a certain number of DPPC molecules are placed randomly into a small box, and then the small box of lipids is centered in a big box of size 20×20×20 nm$^3$ to form a single lipid droplet under vacuum (see Table I). An energy minimization of the system using the method of steepest descent and a long run with a leap-frog stochastic dynamics integrator[58] is carried out for the equilibration at 300K (a *NVT* ensemble for 800 ns effective time) with a time step of 160 fs (effective time, all following scales in time are in effective time.). After the first step, DPPC small balls are centered in the periodic box and the coarse-grained water particles fill the box. To avoid filling of the water into the inside of the balls, we remove the water molecules with a distance of less than 0.5 nm to the closest DPPC molecule. After energy minimization, a position-restrained simulation of DPPC molecules acts on the systems at the corresponding temperature of 300K for 400 ps, i.e., the DPPC molecules are frozen by external forces, and only the water particles can move freely, so that the water can fill up the gap generated by program.

After the position-restrained simulation, the main simulation starts. Twelve total group simulations are carried out, and the number of DPPC molecules in a lipid ball ranges from 100 to 576 (Table I). Twenty parallel samples in each group, which are under the same conditions but have a randomly created initial aggregation structure, are performed. In simulation, the Berendsen thermostat[59], with a coupling time constant of 0.4 ps and a Berendsen pressostat for isotropic pressure coupling are used; the time constant is set to 2.0 ps and the compressibility is 4.5e-5. The time step is set to 160 fs and the total simulation time is 800 ns. All simulations are finished with GROMACS (V3.31).[60]

## 3. RESULTS AND DISCUSSION

The structure of the initial lipid droplet obtained from randomly distributed lipids in a vacuum is not fully disordered. The structure is multilayered (see Fig. 2a, a sample equilibrium state of an initial ball after 800 ns) due to strong intermolecular interactions between the same groups. Fig. 2b and 2c show the formation scheme of these multilayered structures. There are two types of DPPC bilayers in a lipid droplet: the tail-tail bilayer and the head-head bilayer (corresponding to zone 1 and zone 2, respectively, in Fig 2a). In the tail-tail type bilayer, DPPCs' tail groups (hydrophobic parts) join together in the inner part of the layer, and the head groups are located on the outside. When the lipid ball is thrown into water, lipid molecules in the edge of zone 1 spontaneously flip-flop to form a cap (Fig. 3), which decreases the contact of lipid tails with water, because of the hydrophobicity of DPPC tails and the hydrophilicity of DPPC heads. Meanwhile, in zone 2 (head-head type bilayer), the arrangement of DPPCs is the reverse of that in zone 1. The head-head bilayer is torn apart at the edge, with the flip-flop of the DPPC in the opposite direction, and creates a channel where water can intrude into the interior of the lipid ball (Fig. 4b-4d). As a result, the porous structures in the outer layer of the droplet are formed by the following mechanism: lipids in Zone 1 form the wall to stabilize the channel, and the molecules in Zone 2 create the hollow channels to provide space for water to invade.

The formation of water channels in a vesiculation process was also observed in Marrink's early simulations,[41] in which the vesiculation went through several steps: at first, micelles came out from the high concentration zone of the solution, then larger lamella appeared through the coalescence and growth of small micelles; nanoseconds later, many water pores (channels) were observed on the body of a vesicle-like bilayer, where these channels gave lipids a way to equilibrate the inner and outer parts of the bilayer through diffusion movements; after a relatively long time of pore collapse, the vesicle finally formed. The channels in our simulation have a similar function in the transport process of lipids and water, but the formation of these channels is not the same: In Marrink's simulation, the water channels are from the low-density region of the solution. Essentially, they have existed since the beginning of the evolution and influence the whole process of vesiculation. In our system, however, there is no channel in the starting state, and our water channels are formed due to the water-lipids interaction, which gradually affects the arrangement of molecules from the outside to the inside of the lipid ball. So in our simulation, a special swollen lipid ball can be observed in the first several nanoseconds. In addition, the relaxation time of bilayer structures during the coalescence of water pores is much shorter in our simulation as compared to Marrink's process, due to a lack of micelle growth in this system.

With increasing number of DPPCs in the lipid ball, three different final morphologies: spherical micelle, disk-like lamella, and vesicle, are obtained in the final equilibrium stage (see Table I). The trend of morphology transition, i.e., from spherical micelle to vesicle, is in accord with experimental facts.[16,21,61] When the initial droplet is small, only spherical micelles are obtained (group 1). When the number of total lipids increases to 216 (group 5), an interesting splitting phenomenon in the morphology starts to appear (group 5-11 in Table I), i.e., for a DPPC droplet with the same amount of lipid molecules, two fully different equilibrium structures can be obtained. When the size goes up to 576 (group 12), this splitting phenomenon disappears again, but the equilibrium morphology changes to only vesicles. In order to explain this phenomenon, we select two representative samples (S1 and S2 in Fig. 5) from group 9. In the dynamical process, lipids on the multilayered structure rearrange violently in the first several nanoseconds until two similar cuplike intermediates are observed at about 104 ns (see images of C1 and C2 in Fig. 5). Then, the splitting phenomenon appears, i.e., S1 forms a disc and S2 forms a vesicle. We test the stability of the vesicle or the disk structure through a further, longer runtime (8000 ns) (The vesicle and disc structures come from group 5, with 216 lipids, and group 11, with 512 lipids, respectively corresponding to the lower limit of vesicle formation and upper limit of disk formation). In this test, we did not observe any reversible disc-vesicle transition. This means that the final structures are highly stable at this temperature. We extract the evolution energy data of all samples (see supported data), which is calculated as the sum of potential energy and kinetic energy. In the beginning stage, the decrease of the total system energy is associated with an increase in the lipid energy, and is dominated by the interaction between lipids and water. Then, the energy remains constant with a thermal fluctuation. In some samples, a small peak in the lipid energy curve (corresponding to a small plateau in the total energy curve) can be observed regardless of whether the final structure is vesicle or disk. The peak observed for the disk structure is more obvious than that for the vesicle and its appearance frequency increases with increasing lipid droplet size. As an example, Fig. 6 shows the energy changes of the samples S1 and S2 from Fig. 5. After a rapid increase in lipid energy, a weak energy peak (S1) on the DPPC's energy curve appears at about 104 ns, and a small plateau can be also observed in the total energy curve of the system. However, S2 has no such signal at the same time. The peak or platform in the energy curves corresponds to the relaxation and rearrangement process of molecules, after a quick self-assembly in the early stage.

The simulation results show that there are two energy minima, corresponding to disk and vesicle structures through a common intermediate-cuplike structure. The intermediate structure is the key factor in understanding the hidden mechanism. This cuplike structure had been reported in previous studies[20-22,39], where it was regarded as the transient morphology in the process of lamella-vesicle transformation. The case in our simulation, however, is different because our initial structure is a lipid droplet in water (a state with very high energy), without the process of micelle growth and coalescence. In some samples, although the cuplike intermediates are very similar, the evolutions thereafter are totally different: some become disks and some become vesicles. The cup is the intermediate shape between the disc and vesicle: when the cup encloses it becomes a vesicle, when the cup stretches it turns into a disc. Combining the geometry and energy perspective, we propose an ideal hybrid model, based on Helfrich's membrane elastic theory, and try to use this model to explain our simulation and the hidden mechanism behind the simulation results.

Information about the vesicle structural dependence on the size is also gained from a series of simulations. Fig. 7 shows the number of lipids in the inner and outer layer for various equilibrium vesicles. The linear relationship between the two curves is clearly shown. The larger slope for the outer layer curve is due to the relative incompactness of lipid molecules in the outer layer. The radial distribution function (RDF) of the three groups (head, end and tail groups) of lipid molecules relative to the mass center of the vesicle further confirmed this reason (Fig. 8a, extract from the vesicle sample with 392 lipids we discussed above). For the head group curve (red line), the asymmetry density distribution in the two layers indicates the different packing density of lipid molecules; the inner layer possesses less space but a larger density can be observed. The ending group distribution (black line, two end groups of the hydrophobic tail) shows the interface of the two layers. We define the radius of the vesicle to be the peak position of the distribution curve of the end groups. The average surface area of the vesicle sphere at different lipid number is shown in Fig. 8b. The better linear relation provides us with a way to estimate the vesicle size by counting the lipid number. The average lipid number per unit area, $n_a$, is obtained by fitting, i.e., $n_a$= 3.143.

To understand the phenomena of splitting morphologies, we applied a hybrid model based on Helfrich's membrane elastic theory to explain how the cuplike intermediate can bridge the two different final shapes. The basic idea is to find the relationship between the three different morphologies (Disk, Cup and Vesicle). The membrane curvature free energy can be expressed as follows[23, 24]:

$$\begin{cases} g = \frac{1}{2}k_c(c_1+c_2-c_0)^2 + \frac{1}{2}\overline{k}_c c_1 c_2 \\ E_{elas} = \oint_A g dA \end{cases} \quad (1)$$

where $g$ is the bending elastic energy per unit area of the membrane, the total elastic energy $E_{elas}$ is calculated by integrating all units on the area $A$, $c_1$ and $c_2$ are the two principal curvatures of the specified surface, $k_c$ and $\overline{k}_c$ are elastic moduli, and $c_0$ is the spontaneous curvature that is caused by the asymmetric environment. The free energy term is determined by both the geometric structure of the membrane and its surrounding environment. To try to gain a concise expression, some assumptions are applied:[24,52,53] (1) There are only two final morphologies: Lamella and Vesicle. The DPPC bilayer is viewed as an elastic membrane that can be described using Helfrich's membrane elastic theory, and the thickness of the membrane is not considered; (2) The intermediate between the lamella and the vesicle is part of a spherical surface, and the total surface area is a constant value (corresponding to the number of DPPC molecules); (3) Only two energy terms are of interest: The membrane's curvature energy and its free rim energy; (4) The spherical membrane structure is symmetrical, i.e., $c_1=c_2$ and $c_0=0$. So, the unit area elastic energy can be simplified as: $g = 2k_c c^2 + \overline{k}_c c^2 = k'c^2$, where $c$ is curvature. The total energy includes two terms: $E_{elas}$ and $E_{rim}$, i.e., $E_{tot} = E_{elas} + E_{rim}$ where $E_{rim}$ is the free energy of the rim, related to the rim length $L_{rim}$ and energy is constant per unit length $e_{rim}$. So $E_{rim} = L_{rim} e_{rim}$.

The dependence on angle $α$ of the morphology transformation from lamella to vesicle is shown in Fig. 9 (section view of a spherically symmetrical lamellar model). The curvature radii $R$ and curvature $c$ can be expressed as

$$R = \sqrt{\frac{A}{2\pi(1-\cos\alpha)}} \quad \text{and} \quad c = \frac{1}{R} \tag{2}$$

Combining eq. 1, eq. 2, and the total energy of the system, we obtain:

$$\begin{aligned} E_{tot} &= E_{elas} + E_{rim} \\ &= k'c^2 A + L_{rim} e_{rim} \\ &= 2k'\pi(1-\cos\alpha) + \sqrt{\frac{2\pi A}{1-\cos\alpha}} e_{rim} \sin\alpha \\ &= 4k'\pi \sin^2\frac{\alpha}{2} + 2\sqrt{\pi A} e_{rim} \cos\frac{\alpha}{2} \end{aligned} \tag{3}$$

where $k'$ and $e_{rim}$ are assumed to remain constant, independent of the changing environment. In this way, the total energy is only a function of the angle $\alpha$ and the area $A$. Fig. 10a shows the relationship between $E_{tot}$ and $\alpha$ for various surface areas $A$, ranging from 1 to 50 (in units of $k'$ and $e_{rim}$). The left side represents disk-like lamella ($\alpha=0$) and the right represents vesicles ($\alpha = \pi$). With increasing surface area, the energy of the disk also increases, but the vesicle energy is always constant. Each point on an energy curve corresponds to a specified morphology, and the one at the energy peak represents the critical point of the evolution (cuplike intermediate); from this point, the system falls into one of two states (lamella/vesicle) with locally minimal energies (Fig. 10b). The value of $\alpha$ at the peak point of energy (black dot in Fig. 10a), corresponding to the cuplike intermediate state, decreases from $\pi$ to 0 with increasing lamella size (lipid droplet size in our system). This means that the cuplike structure, as an intermediate state, changes from vesicle to lamella. For a larger lipid droplet, the intermediate state shape is close to a lamella, with a small curvature. But for small lipid droplets, the intermediate state is closer to a vesicle-like structure, with large curvature.

Probability theory can be used to explain the relationship between the simulation results and our model. Although two evolutions (lamella and vesicle) through a cuplike intermediate were observed, we can't exactly predict the evolution pathways, because the initial lipid balls are formed randomly. However, we can obtain the probability of each direction, which strongly depends on the equilibrium state energy and the shape of the energy curve. The probability of forming a vesicle increases with the surface area (Fig. 10a). This is in accordance with our simulation results (Table 1). Furthermore, at a certain point ($A=50$ with unit of $k'$ and $e_{rim}$), the probability of forming a lamella is nearly zero, and the vesicle morphology is the single stable state. This result indicates that in the lipid-water system, if the isolated system of a lipid ball or aggregation body is large enough, we will only obtain vesicles instead of disk-like micelles. In a suitable region, the disk and vesicle can coexist. Fig. 11b shows the dependence on droplet size of the energy difference between disk and vesicle structures. When the droplet is small, the energy of the vesicle is higher than that of the disk. When the droplet is larger, the energy of disk is larger than that of vesicle. When the droplet size approaches 360~390, the energies of the disk and the vesicle are almost the same.

From this simple model, some important information can be predicted. By calculating the derivative of the energy with respect to $\alpha$, the eq. 3 becomes:

$$\begin{aligned}\frac{dE_{tot}}{d\alpha} &= 4k'\pi\sin\frac{\alpha}{2}\cos\frac{\alpha}{2} - \sqrt{\pi A}e_{rim}\sin\frac{\alpha}{2} \\ &= \sin\frac{\alpha}{2}(4k'\pi\cos\frac{\alpha}{2} - \sqrt{\pi A}e_{rim})\end{aligned} \quad (4)$$

The peak points of the energy (saddle points) can be obtained by setting the left side of eq. 4 to zero. There must be at least one zero term in the two multiplied terms, i.e.

$$\alpha = 0 \quad \text{or} \quad \alpha = 2\arccos\frac{\sqrt{\pi A}e_{rim}}{4k'\pi} \quad (5)$$

where the second zero term corresponds to the peak energy point. The critical lamella area is obtained from the second term by setting $\alpha$ to 0:

$$A_c = \frac{16k'^2\pi}{e_{rim}^2} \quad (6)$$

When the disk area is above this critical value, the vesiculation trend is so strong that any initial lamella would spontaneously curve into a vesicle. When the disk area is less than this limiting value, the lamella and vesicle can coexist[21, 62]. The probability distribution of the formation of a disk or a vesicle can be analyzed with two state models. Table I shows the final morphologies with different lipid number and different initial lipid sizes. According to the state distribution in thermodynamics, the existence probability of a structure is related to the energy by $e^{-E_P/kT}$. So in our system, the ratio of vesicle appearance can be written as:

$$R_V = \frac{e^{-\frac{E_V}{k_BT}}}{e^{-\frac{E_V}{k_BT}} + e^{-\frac{E_L}{k_BT}}} \quad (7)$$

where $E_V$ and $E_L$ are the structural energy of a vesicle and a lamella, respectively, and the exponential terms represent their formation probability (or the density at a certain point). The $R_V$ provides a vesicle/total-samples ratio in macroscopic observation. Combining eq. 3 and eq. 7, the $R_V$ in our model can be expressed as:

$$R_V = \frac{1}{1+e^{\frac{4k'\pi - 2\sqrt{\pi A}e_{rim}}{k_B T}}} \quad (8)$$

Fig. 11a shows the ratio of vesiculation in the simulation (dotted plot, data from Table 1) and the theoretical model (from eq. 8, best fitting curve). The point of equal vesicle/disk probability is about $A_{0.5}= 388$, which is close to the prediction value (362) from the energy difference, in Fig. 11b. We assume that the area $A$ is proportional to the number of the lipid molecules. So two basic parameters (elastic modulus and rim energy per unit area) in the model are obtained from the fitting method: $k'$=3018 KJ/mol and $e_{rim}$ = 543 KJ/mol·m. The critical area, $A_C$ (the smallest number of DPPCs at 300K) can be calculated from eq. 6, and $A_C$=1553. From the previous section, we obtained the lipid number per unit area: $n_a$=3.143, so the critical vesicle radius can be calculated to be $r_C$=19.7 nm. This predicts that when the number of DPPCs in a lipid ball is below 1553, the splitting phenomena can be observed, and above this value only vesicles can be obtained. It is worthy to note that the model proposed here can be applied not only to lipid systems but also to other amphiphilic systems, such as a surfactant or amphiphilic copolymer system with narrow distribution. Based on our study of vesicle formation from an initial lipid ball, a new method to control vesicle size may be proposed:[64] First, we can prepare small lipid balls with a micro-sprayer using lipid solution. A solvent with low boiling point can be easily evaporated to create lipid balls. The size of the lipid ball can be controlled by tuning the droplet size of sprayed solution and the concentration of the lipid in the solution. Then, lipid balls are thrown into water, and vesicles with controlled size can be obtained through self-assembly of molecules only if the lipid ball is larger than the critical size.

## 4. CONCLUSION

We use Marrink's coarse-grained model to investigate the vesiculation of lipid droplets (DPPC balls). Three equilibrium morphologies (Spherical micelle, Disk-like Lamella, and Vesicle) are at various DPPC droplet sizes. It is discovered that the vesiculation of a lipid ball proceeds from two types of self-assembled patterns of molecules in tail-tail and head-head zones. The special high-energy morphology (Cuplike intermediate) is also observed and an interesting phenomenon of morphology splitting is discovered when the size of DPPC ball is located within a certain range. A hybrid model based on Helfrich's membrane theory is applied, and explains our simulation results well. The important critical parameters, such as the elastic curvature modulus, critical size of lamella, and rim energy are successfully predicted. This method provides us with a bridge between the microscopic events and macroscopic experimental facts. The controllability of vesicle formation and possible application in practice is also discussed.

## 5. ACKNOWLEDGMENT

The author thanks L. Wang's helpful discussion. The financial support of National Science Funds of China (No. 20804028) and RFDP are gratefully acknowledged.

# 6. REFERENCE

[1] M. Almgren, Biochim. Biophys. Acta 1508, 146 (2000).

[2] Suzana Šegota and D¯urd¯ica Težak, Adv. Colloid Interface Sci. 121, 51 (2006).

[3] G. Cevc, Adv. Drug Delivery Rev. 56, 675 (2004).

[4] G. M. El Maghraby, B. W. Barry, and A. C. Williams, Eur. J. Pharm. Sci. 34, 203 (2008).

[5] M. P. Nieh, J. Katsaras, and X. Qi, Biochim. Biophys. Acta - Biomembranes 1778, 1467 (2008).

[6] P. Walde and S. Ichikawa, Biomol. Eng. 18, 143 (2001).

[7] K. Bryskhe, S. Bulut, and U. Olsson, J. Phys. Chem. B 109, 9265 (2005).

[8] A. Veronese, N. Berclaz, and P. L. Luisi, J. Phys. Chem. B 102, 7078 (1998).

[9] X. Luo, W. Miao, S. Wu, and Y. Liang, Langmuir 18, 9611 (2002).

[10] J. Hao, J. Wang, W. Liu, R. Abdel-Rahem, and H. Hoffmann, J. Phys. Chem. B 108, 1168 (2004).

[11] H. Kawasaki, M. Souda, S. Tanaka, N. Nemoto, G. Karlsson, M. Almgren, and H. Maeda, J. Phys. Chem. B 106, 1524 (2002).

[12] B. J. Reynwar, G. Illya, V. A. Harmandaris, M. M. Müller, K. Kremer, and M. Deserno, Nature 447, 461 (2007).

[13] L. Courbin, J. P. Delville, J. Rouch, and P. Panizza, Phys. Rev. Lett. 89, 148305 (2002).

[14] T. Umeda, Y. Suezaki, K. Takiguchi, and H. Hotani, Phys. Rev. E 71, 11913 (2005).

[15] N. Rodriguez, S. Cribier, and F. Pincet, Phys. Rev. E 74, 61902 (2006).

[16] A. J. O'Connor, T. A. Hatton, and A. Bose, Langmuir 13, 6931 (1997).

[17] M. Gradzielski, Curr. Opin. Colloid Interface Sci. 9, 256 (2004).

[18] C. Hamai, P. S. Cremer, and S. M. Musser, Biophys. J. 92, 1988 (2007).

[19] J. Leng, S. U. Egelhaaf, and M. E. Cates, Biophys. J. 85, 1624 (2003).

[20] S. U. Egelhaaf and P. Schurtenberger, Phys. Rev. Lett. 82, 2804 (1999).

[21] A. Shioi and T. A. Hatton, Langmuir 18, 7341 (2002).

[22] M. Ollivon, S. Lesieur, C Grabielle-Madelmont, and M. Paternostre, Biochim. Biophys. Acta - Biomembranes 1508, 34 (2000).

[23] R. Capovilla, J. Guven, and J. A. Santiago, Phys. Rev. E 66, 021607 (2002).

[24] Z. C. Tu and Z. C. Ou-Yang, Phys. Rev. E 68, 061915 (2003).

[25] G. Battaglia and A. J. Ryan, Nat. Mater. 4, 869 (2005).

[26] T. M. Weiss, T. Narayanan, and M. Gradzielski, Langmuir 24, 3759 (2008).

[27] T. M. Weiss, T. Narayanan, C. Wolf, M. Gradzielski, P. Panine, S. Finet, and W. I. Helsby, Phys. Rev. Lett. 94, 38303 (2005).

[28] H. -G. Döbereiner, E. Evans, M. Kraus, U. Seifert, and M. Wortis, Phys. Rev. E 55, 4458 (1997).

[29] M. Venturoli, M. M. Sperotto, M. Kranenburg, and B. Smit, Phys. Rep. 437, 1 (2006).

[30] H. Noguchi and M. Takasu, Phys. Rev. E 64, 041913 (2001).

[31] V. Knecht and S. J. Marrink, Biophys. J. 92, 4254 (2007).

[32] S. Yamamoto, Y. Maruyama, and S. Hyodo, J. Chem. Phys. 116, 5842 (2002).

[33] N. Arai, K. Yasuoka, and Y. Masubuchi, J. Chem. Phys. 126, 244905 (2007).

[34] J. M. Drouffe, A. C. Maggs, and S. Leibler, Science 254, 1353 (1991).

[35] D. J. Michel and D. J. Cleaver, J. Chem. Phys. 126, 034506 (2007).

[36] J. Marrink, D. P. Tieleman, and A. E. Mark, J. Phys. Chem. B 104, 12165 (2000).

[37] H. Noguchi and G. Gompper, J. Chem. Phys. 125, 164908 (2006).

[38] P. Ballone and P. M. Del, Phys. Rev. E 73, 31404 (2006).

[39] S. J. Marrink and A. E. Mark, J. Am. Chem. Soc. 125, 15233 (2003).

[40] A. H. de Vries, A. E. Mark, and S. J. Marrink, J. Am. Chem. Soc. 126, 4488 (2004).

[41] P. M. Kasson, N. W. Kelley, N. Singhal, M. Vrljic, A. T. Brunger, and V. S. Pande, Proc. Natl Acad. Sci. USA 103, 11916 (2006).

[42] S. J. Marrink, J. Risselada, and A. E. Mark, Chem. Phys. Lipids 135, 223 (2005).

[43] B. Y. Wong and R. Faller, Biochim. Biophys. Acta - Biomembranes 1768, 620 (2007).

[44] K. Dimitrievski and B. Kasemo, Langmuir 24, 4077 (2008).

[45] S. Leekumjorn and A. K. Sum, Biochim. Biophys. Acta - Biomembranes 1768, 354 (2007).

[46] M. Orsi, D. Y. Haubertin, W. E. Sanderson, and J. W. Essex, J. Phys. Chem. B 112, 802 (2008).

[47] T. Murtola, E. Falck, M. Patra, M. Karttunen, and I. Vattulainen, J. Chem. Phys. 121, 9156 (2004).

[48] D. Bemporad, J. W. Essex, and C. Luttmann, J. Phys. Chem. B 108, 4875 (2004).

[49] E. R. May, A. Narang, and D. I. Kopelevich, Phys. Rev. E 76, 21913 (2007).

[50] C. F. Lopez, S. O. Nielsen, P. B. Moore, J. C. Shelley, and M. L. Klein, J. Phys.: Condens. Matter 14, 9431 (2002).

[51] W. Helfrich, Z. Naturforsch. C 28, 693 (1973).

[52] D. Ni, H. Shi, and Y. Yin, Colloids Surf. B: Biointerfaces 46, 162 (2005).

[53] S. Lorenzen, R. M. Servuss, and W. Helfrich, Biophys. J. 50, 565 (1986).

[54] L. M. Bergström, J. Colloid Interface Sci. 293, 181 (2006).

[55] S. J. Marrink, H. J. Risselada, S. Yefimov, D. P. Tieleman, and A. H. deVries, J. Phys. Chem. B 111, 7812 (2007).

[56] R. Notman, M. G. Noro, and J. Anwar, J. Phys. Chem. B 111, 12748 (2007).

[57] C. Xing and R. Faller, J. Phys. Chem. B 112, 7086 (2008).

[58] W. F. Van Gunsteren and H. J. C. Berendsen, Mol. Simul. 1, 173 (1988).

[59] H. J. C. Berendsen, J. P. M. Postma, W. F. van Gunsteren, A. DiNola, and J. R. Haak, J. Chem. Phys. 81, 3684 (1984).

[60] D. Van Der Spoel, E Lindahl, B Hess, G Groenhof, A. E. Mark, and H. J. C. Berendsen, J. Comput. Chem. 26, 1701 (2005).

[61] D. D. Lasic, R. Joannic, B. C. Keller, P. M. Frederik, and L. Auvray, Adv. Colloid Interface Sci. 89-90, 337 (2001).

[62] H. -T. Jung, S. Y. Lee, E. W. Kaler, B. Coldren, and J. A. Zasadzinski, Proc. Natl Acad. Sci. USA 99, 15318 (2002).

[63] W. Humphrey, A. Dalke, and K. Schulten, J. Mol. Graph. 14, 33 (1996).

[64] H. S. Bu, E. Q. Chen, S. Y. Xu, K. X. Guo, and B. Wunderlich, J. Polym. Sci. Polym. Phys. Ed. 32, 1351 (1994).

# 7. TABLES

TABLE I. Overview of the DPPC-Water system and the results of the simulation

| Group | DPPC | Equilibrium morphology (ratio) * |
|:-----:|:----:|:--------------------------------:|
| 1     | 100  | SM (20/20)                       |
| 2     | 125  | DL (20/20)                       |
| 3     | 150  | DL (20/20)                       |
| 4     | 180  | DL (20/20)                       |
| 5     | 216  | DL (19/20) and V( 1/20)          |
| 6     | 252  | DL (19/20) and V( 1/20)          |
| 7     | 294  | DL (16/20) and V( 4/20)          |
| 8     | 343  | DL (16/20) and V( 4/20)          |
| 9     | 392  | DL ( 9/20) and V(11/20)          |
| 10    | 448  | DL ( 5/20) and V(15/20)          |
| 11    | 512  | DL ( 2/20) and V(18/20)          |
| 12    | 576  | V(20/20)                         |

* Symbols: SM - Spherical Micelle. DL - Disk-like Lamella. V - Vesicle

## 8. FIGURE CAPTIONS:

Figure. 1   A coarse-grained DPPC model developed by Marrink et al[55]. Four major atoms correspond to one coarse-grained bead.

Figure. 2   Scheme of the aggregation structure of lipids: a: initial DPPC small ball (512 lipids, color online), where solid yellow and dark gray are the head groups and the tail groups respectively; b: randomly distributed DPPC molecules; c: DPPC multilayer structure.

Figure. 3   A scheme of the interaction of DPPC with water in the tail-tail zone: the hydrophilic head groups join together to prohibit the invasion of water.

Figure. 4   A scheme of the interaction of DPPC with water in the head-head zone. The hydrophobic tail groups join together and the head-head zone splits to form a channel for water's pervasion.

Figure. 5   Morphology evolution (middle section) of two samples (S1 and S2, corresponding to disk and vesicle respectively) in group 9 of Table 1 (392 DPPC molecules, water isn't shown, color online). Two CG particles of the head group in DPPC are colored red and blue for distinction, and tail groups are colored in black. Images are rendered using VMD package[63]. (A) 0 ns, (B) 8 ns, (C) 104 ns, (D) 240 ns, (E) 800 ns.

Figure. 6   The dependences on the time of energy changes in two samples (S1 and S2, corresponding to disk and vesicle respectively) in Fig. 5: Upper panel is the energy evolution of the total system (including DPPCs and water); Lower panel represents the energy curve of DPPCs without water. The energy is the sum of kinetic and potential energy. The arrow line with time labeled as 104 ns is the peak energy point corresponding, to the cuplike intermediate structure.

Figure. 7   The average lipid number in the inner and outer layers of an equilibrium vesicle with different vesicle sizes (dot). The linear relationship with the total number of DPPCs is shown by fitting straight lines.

Figure. 8   a: The radial distribution function (RDF) curve of three characteristic groups (head group, end group, and the hydrophobic chain-tail group, color online) with respect to the mass center of the vesicle (392 DPPCs); b: Dependence of the vesicle area on the total amount of DPPC in the vesicle, where the radius is defined as the distance between the center of mass and the peak point of end group distribution in

RDF.

Figure. 9 Illustration of morphology change by lamellar bending. The curved lamella at the origin point is regarded as a partial spherical surface with curvature radius $R$ and with angle $\alpha$ between the surface edge direction and the minus direction of the Z axis. Figure shows a slice along the XOZ coordinate plane.

Figure. 10 Upper panel: The curves of $E_{tot}$ with different surface area $A$. Here the dark points label the peak energy of each curve, one of which is in bold to be shown more clearly. The left side represents the lamellar morphology, and the right side represents the vesicle. Lower panel: Illustration of two different evolution pathways from an initial lipid ball.

Figure. 11 The dependences, on the droplet size, of (a) vesicle appearance probabilities $R_V$ and (b) the energy difference between disks and vesicles. The droplet size, which is equal to about 362 when the energy of a vesicle is equal to that of a disk, is close to the value (about 388) at $R_V=0.5$ (crossing dash lines).

1
2
3
4
5
6
7
8
9
10
11
12
13
14
15

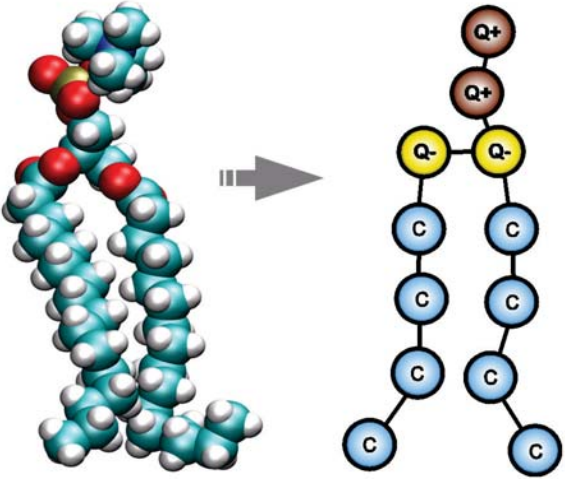

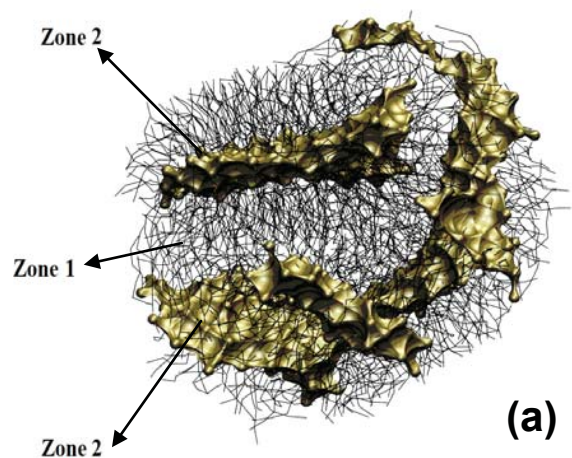
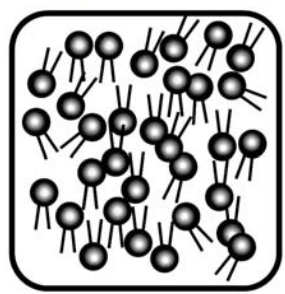
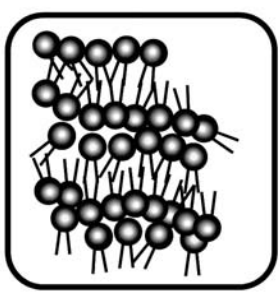

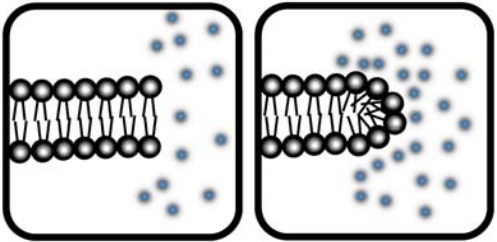
(a)    (b)

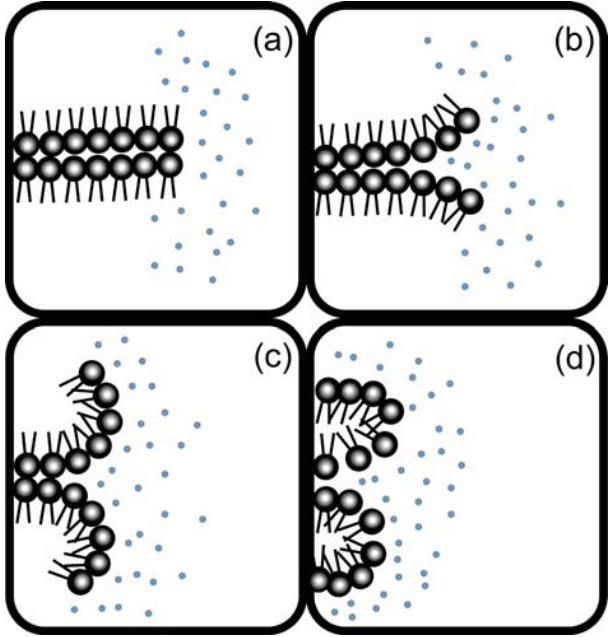

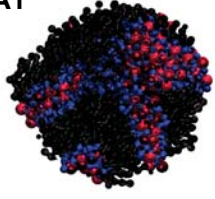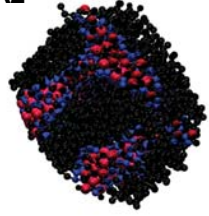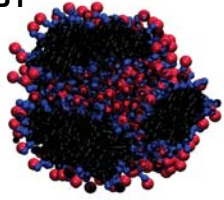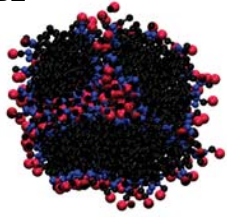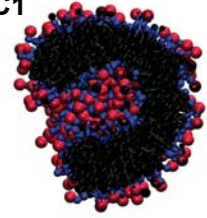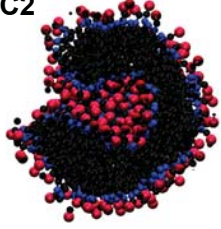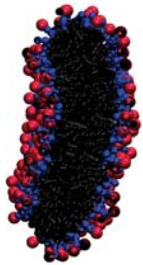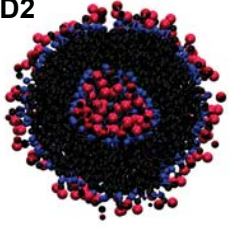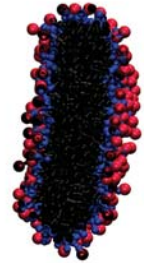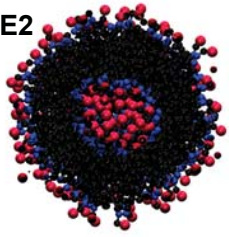

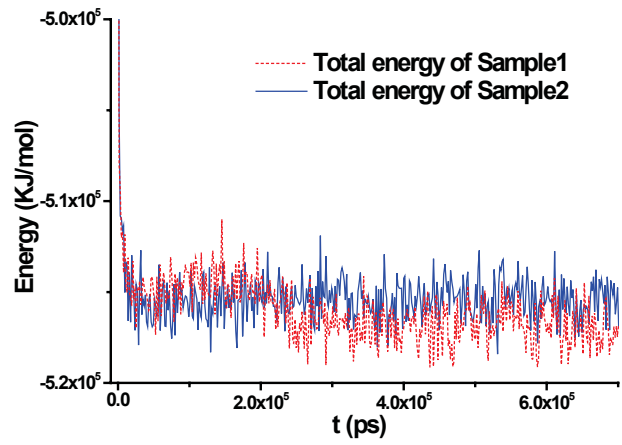

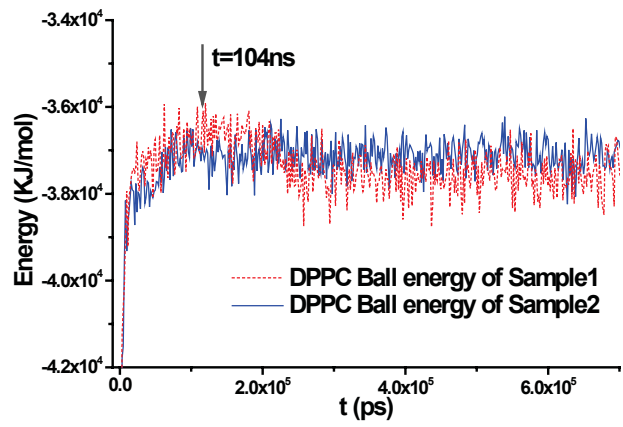

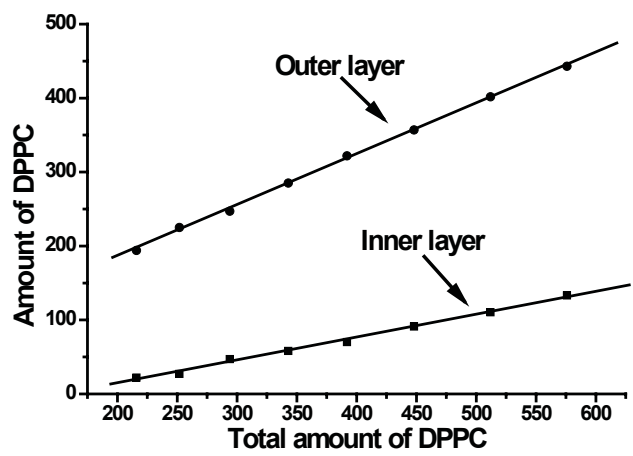

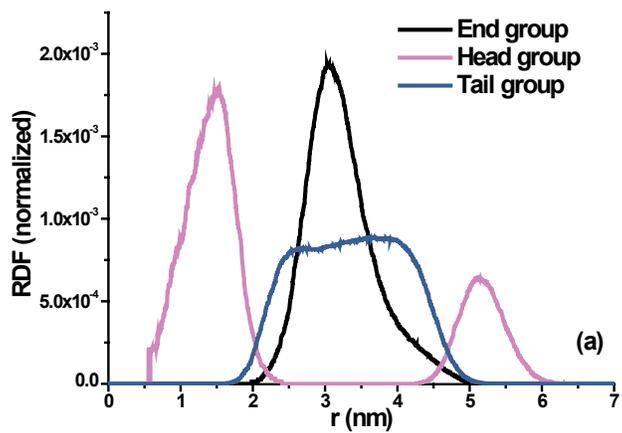

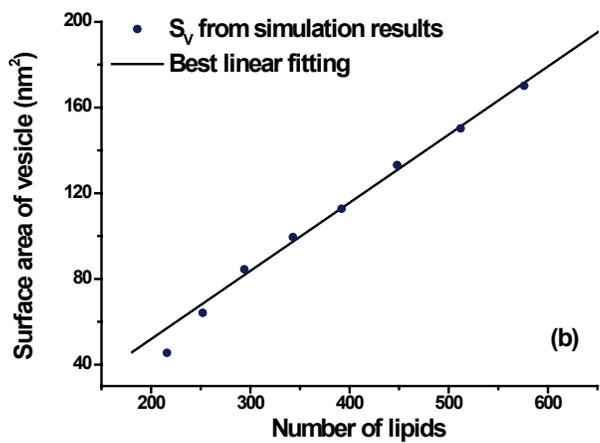

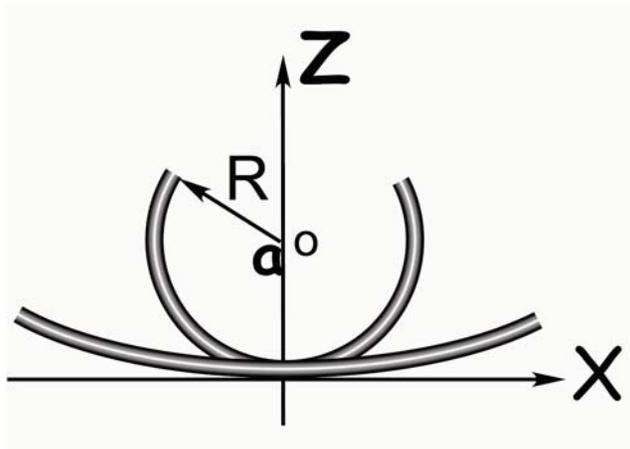

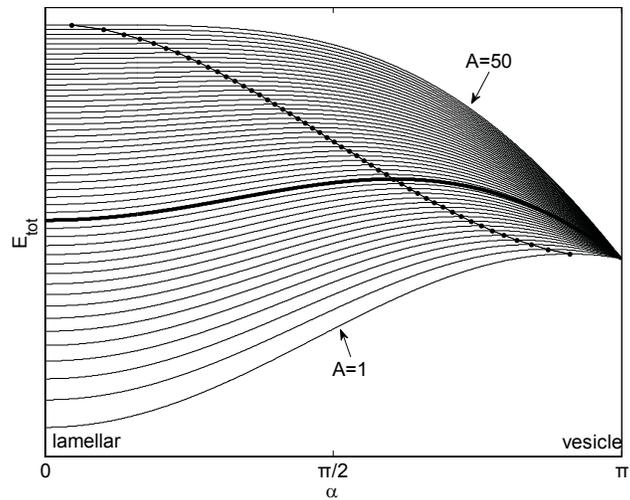

**(a)**

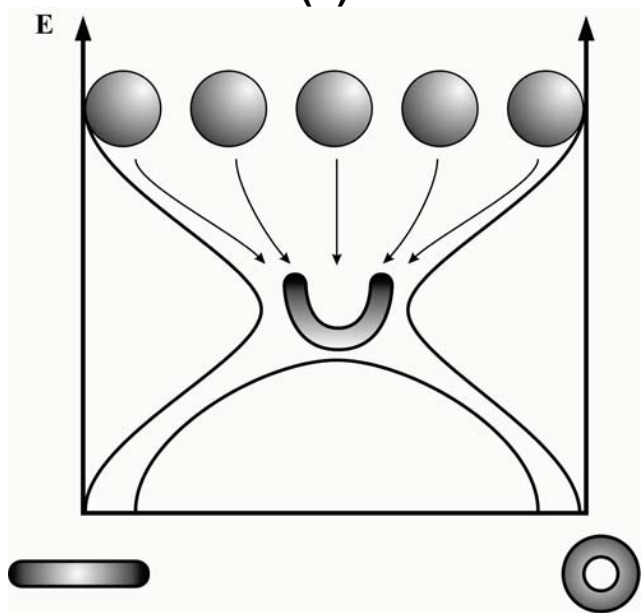

**(b)**

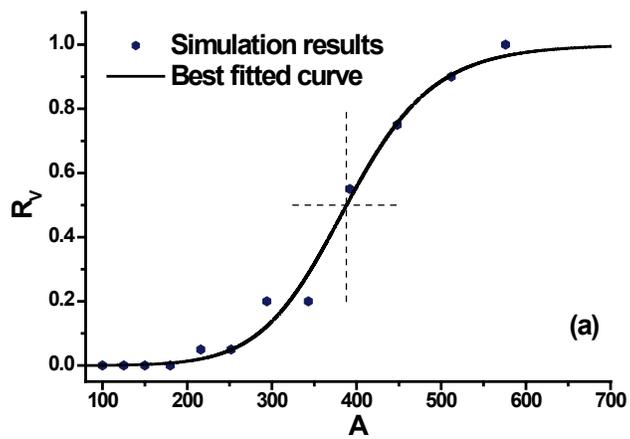
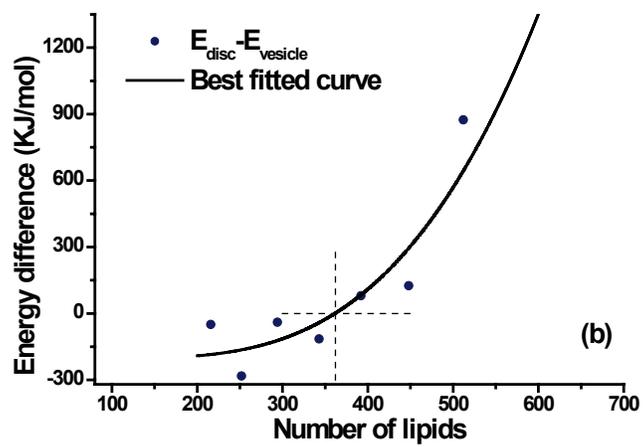